\documentclass[twocolumn]{IEEEtran}

\usepackage[pdftex]{graphicx} 
\usepackage{amsmath} 
\usepackage{algorithmic}
\usepackage{array} 
\usepackage[caption=false,font=footnotesize]{subfig} 
\usepackage{dblfloatfix}
\usepackage{url} 

\usepackage{microtype}
\usepackage[utf8]{inputenc} 
\usepackage[T1]{fontenc}    
\usepackage{paralist} 
\usepackage{marvosym} 
\usepackage{bm} 
\usepackage{bbm} 
\usepackage{amssymb}   
\usepackage{mathrsfs} 
\usepackage{tabularray} 
\UseTblrLibrary{booktabs}       
\usepackage{tabularx} 
\usepackage[flushleft]{threeparttable} 
\usepackage{colortbl} 
\usepackage{makecell} 
\usepackage{multirow} 
\usepackage{nicefrac} 
\usepackage[dvipsnames]{xcolor} 
\usepackage{soul} 
\setul{0.5ex}{0.3ex}

\definecolor{citecolor}{RGB}{34,139,34}
\usepackage[pagebackref=true,breaklinks=true,colorlinks,
    citecolor=citecolor,bookmarks=false]{hyperref}
\usepackage{cleveref}  
\crefformat{figure}{Fig.~#2#1#3}
\crefformat{equation}{Eq.~(#2#1#3)}
\crefformat{table}{Table~#2#1#3}
\crefformat{section}{Section~#2#1#3}
\crefformat{algorithm}{Algorithm~#2#1#3}
\crefformat{appendix}{Appendix #2#1#3}
\crefformat{subsection}{subsection~#2#1#3}
\usepackage{cellspace}
\usepackage[numbers,sort&compress]{natbib} 

\usepackage{float}
\usepackage{xspace} 

\usepackage{pifont} 
\newcommand{\xmark}{\ding{53}}%

\definecolor{Gray}{rgb}{0.9,0.9,0.9}
\definecolor{LightCyan}{rgb}{0.88,1,1}
\newcolumntype{a}{>{\columncolor{Gray}}c}

\begin{document}

\setlength{\abovedisplayskip}{.5\baselineskip} 
\setlength{\belowdisplayskip}{.5\baselineskip} 

\title{Zoom Out and Zoom In: Re-Parameterized Sparse Local Contrast Networks for Airborne \\ Small Target Detection}




\title{Make Both Ends Meet: A Synergistic Optimization Infrared Small Target Detection with Streamlined Computational Overhead}

\author{

  Yuxin~Jing,
  Yuchen~Zheng,  
  Jufeng~Zhao,
  Guangmang~Cui,
  Tianpei~Zhang.
}



\maketitle


\begin{abstract}

Infrared small target detection(IRSTD) is widely recognized as a challenging task due to the inherent limitations of infrared imaging, including low signal-to-noise ratios, lack of texture details, and complex background interference. While most existing methods model IRSTD as a semantic segmentation task, but they suffer from two critical drawbacks: (1) blurred target boundaries caused by long-distance imaging dispersion; and (2) excessive computational overhead due to indiscriminate feature stackin. To address these issues, we propose the Lightweight Efficiency Infrared Small Target Detection (LE-IRSTD), a lightweight and efficient framework based on YOLOv8-n, with following key innovations. Firstly, we identify that the multiple bottleneck structures within the C2f component of the YOLOv8-n backbone contribute to an increased computational burden. Therefore, we implement the Mobile Inverted Bottleneck Convolution block (MBConvblock) and Bottleneck Structure block (BSblock) in the backbone, effectively balancing the trade-off between computational efficiency and the extraction of deep semantic information. Secondly, we introduce the Attention-based Variable Convolution Stem (AVCStem) structure, substituting the final convolution with Variable Kernel Convolution (VKConv), which allows for adaptive convolutional kernels that can transform into various shapes, facilitating the receptive field for the extraction of targets. Finally, we employ Global Shuffle Convolution (GSConv) to shuffle the channel dimension features obtained from different convolutional approaches, thereby enhancing the robustness and generalization capabilities of our method. Experimental results demonstrate that our LE-IRSTD method achieves compelling results in both accuracy and lightweight performance, outperforming several state-of-the-art deep learning methods. The source codes are available at https://github.com/jing2024star/LE-IRSTD.


\end{abstract}

\begin{IEEEkeywords}
IRSTD, lightwight, Mobile Inverted Bottleneck Convolution, Bottleneck Structure Block, Attention-based Variable Convolution Stem, Variable Kernel Convolution, Ghost Shuffle Convolution, YOLO (You only look once)
\end{IEEEkeywords}
\vspace{-1\baselineskip}

\section{Introduction} \label{sec:introduction}
%
Infrared imaging technology is widely used in nighttime surveillance \cite{wu2023mtu,hu2024smpisd} and military reconnaissance \cite{deng2016small} due to its ability to capture thermal radiation. As a core component of infrared imaging and warning systems, infrared small target detection (IRSTD) faces critical challenges in modern applications such as missile defense \cite{tidrow2001infrared} and border monitoring \cite{bagavathiappan2013infrared}. 

\begin{figure}
    \centering
    \includegraphics[width=1.0\linewidth]{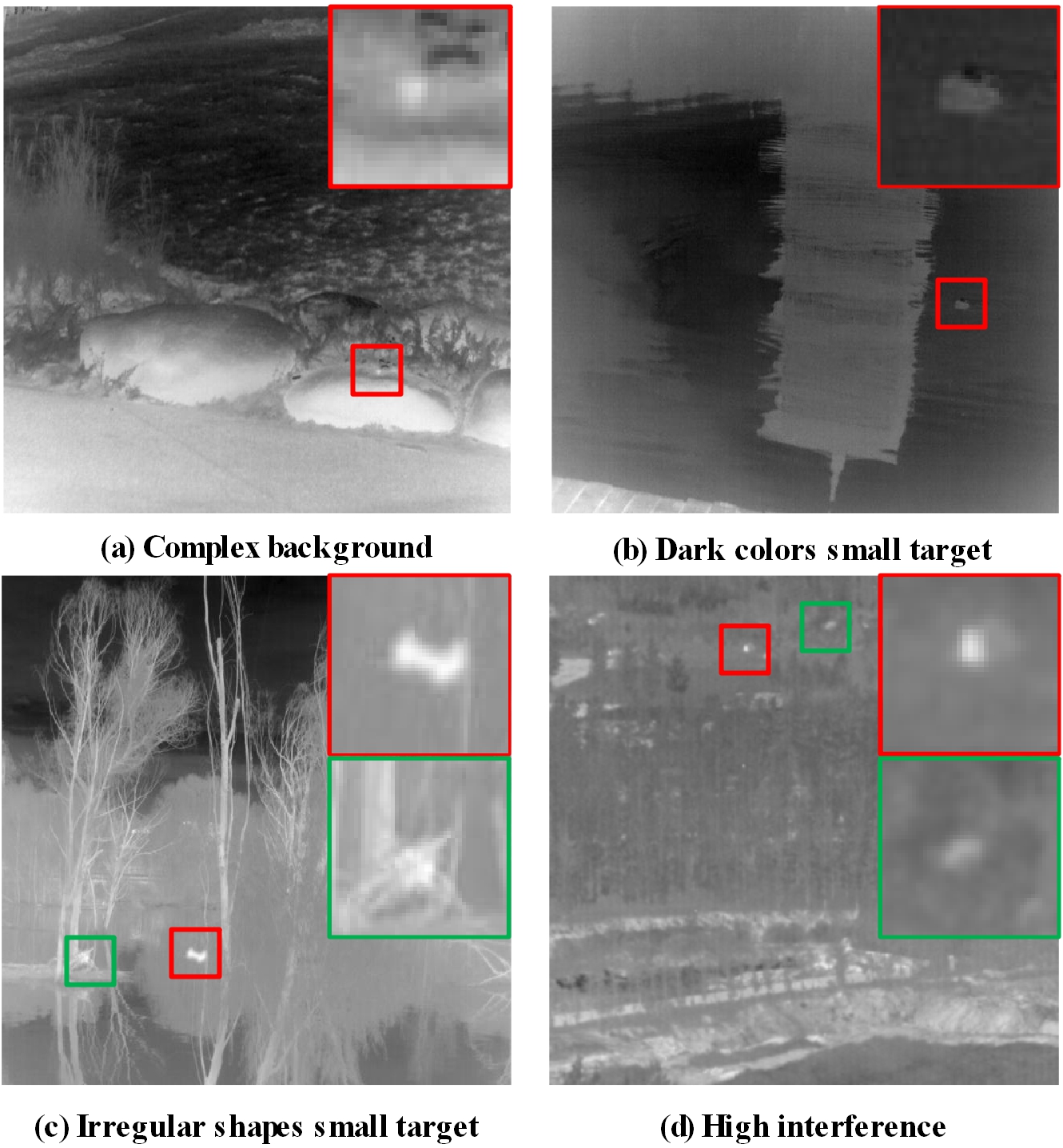}
    \caption{The different small targets exist in the infrared image, The \textcolor{red}{red box} represents a small target, and the \textcolor{green}{green box} represents a false alarm: (a) on complex background, the small target is not obvious. (b) the color of the small target is dim. (c) the shape of the small targeti is irregular. (d) the target is small, and there is more interference}
    \label{fig:fig1background}
\end{figure}


As shown in Fig. \ref{fig:fig1background}, infrared small targets often lack distinct texture and shape features in the image (\textcolor{red}{red bounding box}) and are susceptible to similar interference from the background (\textcolor{green}{green bounding box}). Due to the inherent characteristics of infrared imaging-low resolution, lack of texture details, and color information. Meanwhile, due to the long imaging distance of small targets, which are easily submerged in noise, leading to high false alarm rates and missed detections. Specifically, IRSTD faces the following key challenges:

\begin{enumerate}

\item{\textbf{Background Interfere}:} High contrast false alarms are prone to occur from the background, making it difficult to identify actual targets and false alarms;

\item{\textbf{Lack of feature information}:} Due to the long imaging distance, small targets lack color and texture information, making feature extraction difficult;

\item{\textbf{Consumption of Resources}:} Efficient IRSTD often relies on complex feature extraction algorithms, which require a significant amount of time and computational resources.

\end{enumerate}



\begin{figure}
    \centering
    \includegraphics[width=1\linewidth]{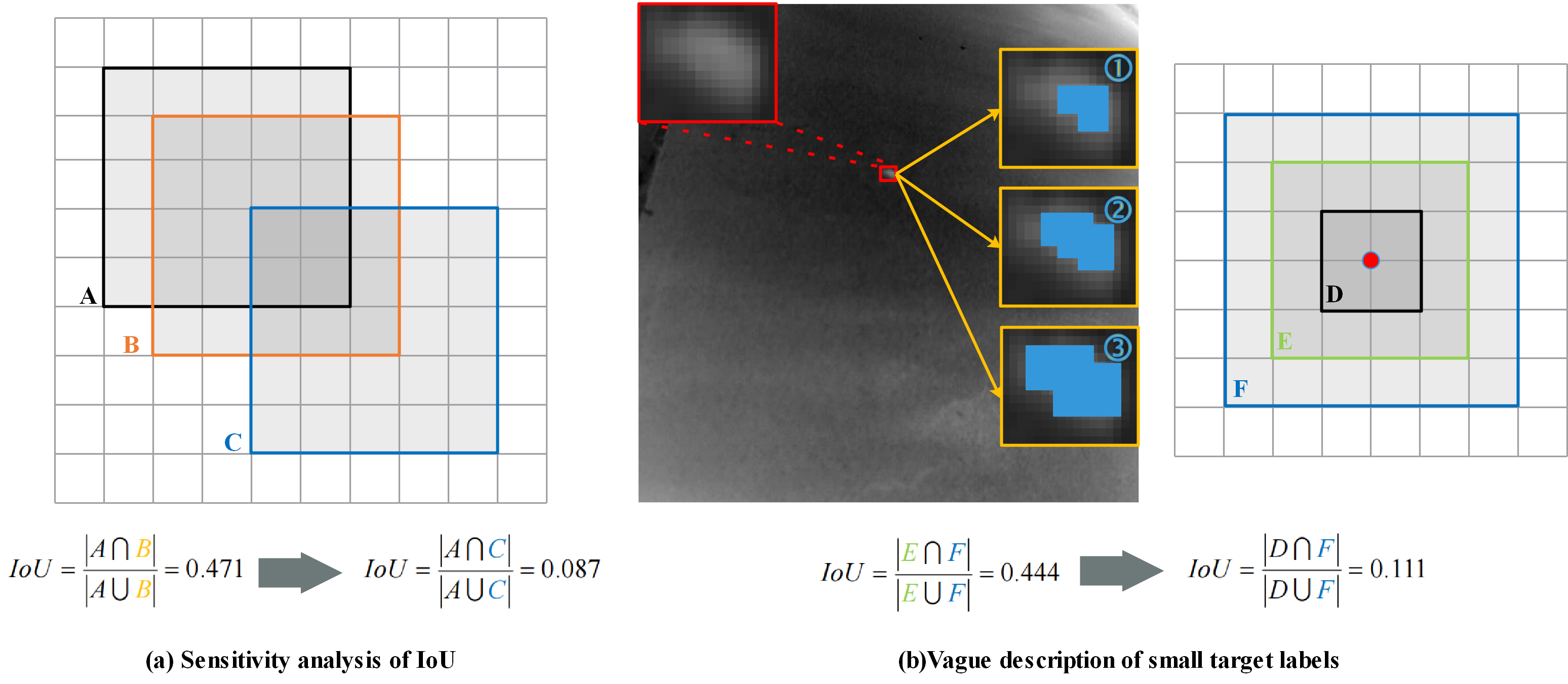}
    \caption{Disadvantages of IoU for small targets. (a) Sensitivity analysis of IoU is illustrated, where box A denotes the actual bounding box, and boxes B and C represent the predicted bounding boxes with diagonal deviations of 1 and 3, respectively. (b) The blue masks 1, 2 and 3 in the left image represent annotations of different ways for small targets, corresponding to the right D, E, F, sharing the same target centroid, boxes D, E represent possible true boxes, and F represents prediction boxes. It can be seen that the small gap between the predicted box and the real box will lead to a significant decrease in IoU.}
    \label{fig:fig2iouanalysis}
\end{figure}

Existing approaches predominantly model IRSTD as a semantic segmentation task, focusing on local information enhancement through edge detection \cite{kou2023infrared, zhu2024towards,zhang2022isnet}, feature gradient analysis \cite{hu2024gradient, sheth2019feature, zhang2025m4net}, and local contrast attention adjustments \cite{dai2021attentional, chen2013local, zhong2024hierarchical}. Despite their progress in IRSTD, it has been difficult to further improve detection performance, even with deeper backbones \cite{zhao2022single} and more complex attention modules \cite{zhao2022single, zhong2024hierarchical}. Specifically, we argue that the fundamental bottleneck of IRSTD lies in the task being modeled as semantic segmentation, which is not optimal. Firstly, \textbf{Task mismatch:} semantic segmentation is only an intermediate representation of small targets, which is merely a rough approximation of the final detection result. Secondly, \textbf{Annotation Ambiguity:} as shown in Fig. \ref{fig:fig2iouanalysis}, the influence of these uncertain pixels on the Intersection over Union (IoU) is extremely sensitive: even a small number of pixel prediction errors can lead to a "cliff drop" in IoU. Meanwhile, due to the dispersion effect of long-distance imaging, making pixel-level annotations inherently unreliable \cite{dai2023one}, which will ultimately limit IoU performance. Thirdly, \textbf{Computational Overhead:} in order to accurately segment infrared small targets, it is necessary to infer high-resolution feature maps, which requires a large amount of computation and is time-consuming. Therefore, considering detection accuracy and computational cost, using bounding boxes to directly detect small targets is more reasonable than relying on segmentation.


While bounding box regression offers computational advantages over segmentation, exiting bounding boxes for IRSTD still face some limitations. Specifically, the current bounding box-based detection improves performance by striking a balance between performance and computational overhead. For instance, YOLO-SDLUWD \cite{zhu2023yolo} increases detection accuracy by introducing a dual-branch structure with an attention mechanism to enhance the local module and by incorporating Vision Transformer \cite{vaswani2017attention} to maximize the expressive power of the feature map. While extracting as much features as possible, this approach significantly increases the computational overhead. Meanwhile, EFLNet \cite{yang2024eflnet} proposed a dynamic head attention mechanism incorporated into the network to achieve adaptive learning of the relative importance of each semantic layer. Overall, existing methods still have have an inherent trade-off delimma:

\begin{enumerate}

\item{\textbf{Large convolutional stacks}:} stacking more convolutional blocks to retain the feature information of small targets will greatly increase the model burden;

\item{\textbf{Semantic information loss}:} the pooling layer operation and the encoding and decoding structure make the high-level and low-level feature details improperly combined, resulting in the feature loss;

\item{\textbf{Complex Computational Overhead}:} large-scale convolutional layers and the introduction of attention mechanisms affect computational overhead.

\end{enumerate}
These difficulties inspire us to achieve "\textbf{make both ends meet}" rather than "\textbf{trade a balance}" between the performance and computational overhead(parameter, Flops). It is necessary to improving the feature extraction capabilities without adding redundant modules to increase Flops and model parameters.

%

Based on the above analysis, the design principles for efficient IRSTD needs to comprehensively consider both feature extraction and a lightweight structure. In feature extraction, an accurate feature extractor with low parameter and Flops is required, as shown in Tab. \ref{tab:sota}, various models have been given the burden of parameter and Flops. In lightweight structure, adopting a streamlined network structure, our model achieves more accurate target positioning and reduces the loss of semantic information. This not only improves detection accuracy but also makes the model more suitable for real-time implementation. Therefore, we need to design a network that aims to overcome false alarms, missed detections, and ensure real-time detection. Unlike conventional attention-based methods \cite{zhao2022single}, experimental validations shows our model significantly reduces computational complexity while maintaining excellent performance.



We proposed lightweight efficient infrared small target detection model (LE-IRSTD), which \textbf{make both ends meet} in lightweight design and detection accuracy. LE-IRSTD derived from YOLOv8-n, which core improvements are in the backbone and neck structures, thereby reducing resource consumption while ensuring high detection accuracy. Firstly, the lightweight backbone replaces standard C2f modules with the MBConvblock and BSblock to better balance the depth, width, and resolution of the network, making the model structure more lightweight. Secondly, the AVCStem module enables cross-stage fusion through variable-kernel convolution (VKConv), which is used to give the convolution kernel an arbitrary number of parameters and sampling shapes to improves the ability to extract feature information. Furthermore, in the PAN structure of the neck, ordinary convolution is replaced by GSConv, thereby reducing the partial loss of semantic information.

The contribution of this paper are as follows:

\begin{enumerate}

\item We designed the obile Inverted Bottleneck Convolution block (MBConvblock) and Bottleneck Structure block (BSblock) as the main feature components of the backbone feature extraction network, optimizing the backbone network by combining network depth, width, and input resolution, while aggregating channel information to effectively capture deep semantic information of target features.

\item We realized the Attention-based Variable Convolution Stem (AVCStem) module of cross-stage feature fusion, and used Variable Kernel Convolution (VKConv) to assign arbitrary number of parameters and sampling shapes to convolution for small targets with different shapes, so as to recognize and fuse targets at different levels.

\item Our proposed Ghost Shuffle Convolution (GSConv) performs operations with different convolution blocks to cross-connect information across the channel dimension, minimizing the loss of semantic information when spatial information is transferred to the channel.
\item The detection head uses classification and regression loss to output and focus the infrared small target with the highest confidence, and three detection heads of different scales output detection results from AVCStem.

\end{enumerate}





\section{Related Work} \label{sec:related}
\subsection{Traditional Methods for IRSTD}

The traditional methods paradigms include filter-based methods, local contrast-based methods, and low-rank decomposition-based methods. Filter-based methods extract small targets by enhancing the difference between the target and the background, such as MaxMedian \cite{deshpande1999max} and Tophat \cite{bai2010analysis} filters. They have shown good performance on smooth or low-frequency backgrounds but exhibit local limitations when dealing with complex backgrounds \cite{liu2023infrared}. Local contrast methods extract targets by enhancing local contrast and suppressing background clutter. For example, local contrast measure (LCM) \cite{chen2013local}, weighted enhanced local contrast measurement (WSLCM) \cite{han2020infrared}, and three-layer local contrast measurement (TLLCM) \cite{han2019local} assume that the brightness of the target is higher than that of its adjacent area, and therefore struggle to detect dim targets. Moreover, low-rank decomposition-based methods use patch tensors for low-rank matrix decomposition. The image is restored through a reverse reconstruction process, including infrared patch images (IPI) \cite{huang2019infrared} and reweighted infrared patch tensors \cite{dai2017reweighted} via rank minimization. Based on the assumption that low-rank backgrounds and sparse targets can separate the target from the background. These methods rely heavily on handcrafted designed features, which can only detect specific features, lacking generalization ability.


\subsection{Deep Learning Methods for IRSTD}

With the development of deep learning methods, as well as the availability of a large number of IRSTD datasets, more and more researchers are becoming increasingly interested in deep learning methods, which can be further divided into segmentation-based methods and detection-based methods.

\subsubsection{\textbf{Segmentation-Based Methods}}

The segmentation-based method uses pixel-by-pixel segmentation to generate segmentation masks that provide object position and size information \cite{zhu2025towards}. For example, DNANet \cite{ren2021dnanet} adopts a densely connected network for multi-level feature fusion and extraction, but this approach increases computational complexity while enhancing the usefulness of target features. ACMNet \cite{qu2021acm} utilizes global attention to help the model understand the relationships between targets, while local attention focuses on fine features to process targets of different scales and shapes, thus improving the robustness of the model. GSTUnet \cite{zhu2024towards} introduces multiple types of supervised signals (such as semantic segmentation maps and edge features) to enable the model to learn target features from different levels, enhancing its generalization ability. However, semantic segmentation has not addressed the essence of IRSTD, which lies in the precise recognition and localization of small targets. Moreover, high accuracy performance corresponds to greater computational costs and higher pixel-level data annotation requirements. These lilmitations highlight the need for IRSTD reformulation beyond semantic segmentation.


\subsubsection{\textbf{Detection-Based Methods}}

The detection-based method is analogous to object detection algorithms and can directly recognize and locate small targets in complex backgrounds \cite{xiao2024background}. To enhance the detection performance of infrared small targets, YOLO-SLWD \cite{zhu2023yolo} improves detection accuracy by introducing a dual-branch structure with attention. The One Stage Cascaded Refinement Network (OSCAR) \cite{dai2023one} is employed to address the issues of insufficient inherent features and inaccurate bounding box regression in infrared small target detection. RetinaNet \cite{ross2017focal} introduces Focal Loss to tackle the problem of imbalanced foreground and background samples, using a feature pyramid network (FPN) to generate multi-scale features. EFLNet \cite{yang2024eflnet} proposes a new adaptive threshold focus loss, normalized Gaussian Wasserstein distance (NWD), and a dynamic head attention mechanism, achieving good results; however, its model complexity is excessively high. The method based on bounding boxes considers IRSTD as an accurate regression problem of bounding boxes, effectively extracting target features while significantly reducing model computational complexity. Therefore, in detection-based methods, it is essential not only to achieve good results but also to maintain models with low complexity and excellent real-time performance.

\section{Method} \label{sec:method}

\subsection{Overall Structure of LE-IRSTD}
\begin{figure*}
    \centering
    \includegraphics[width=0.9\linewidth]{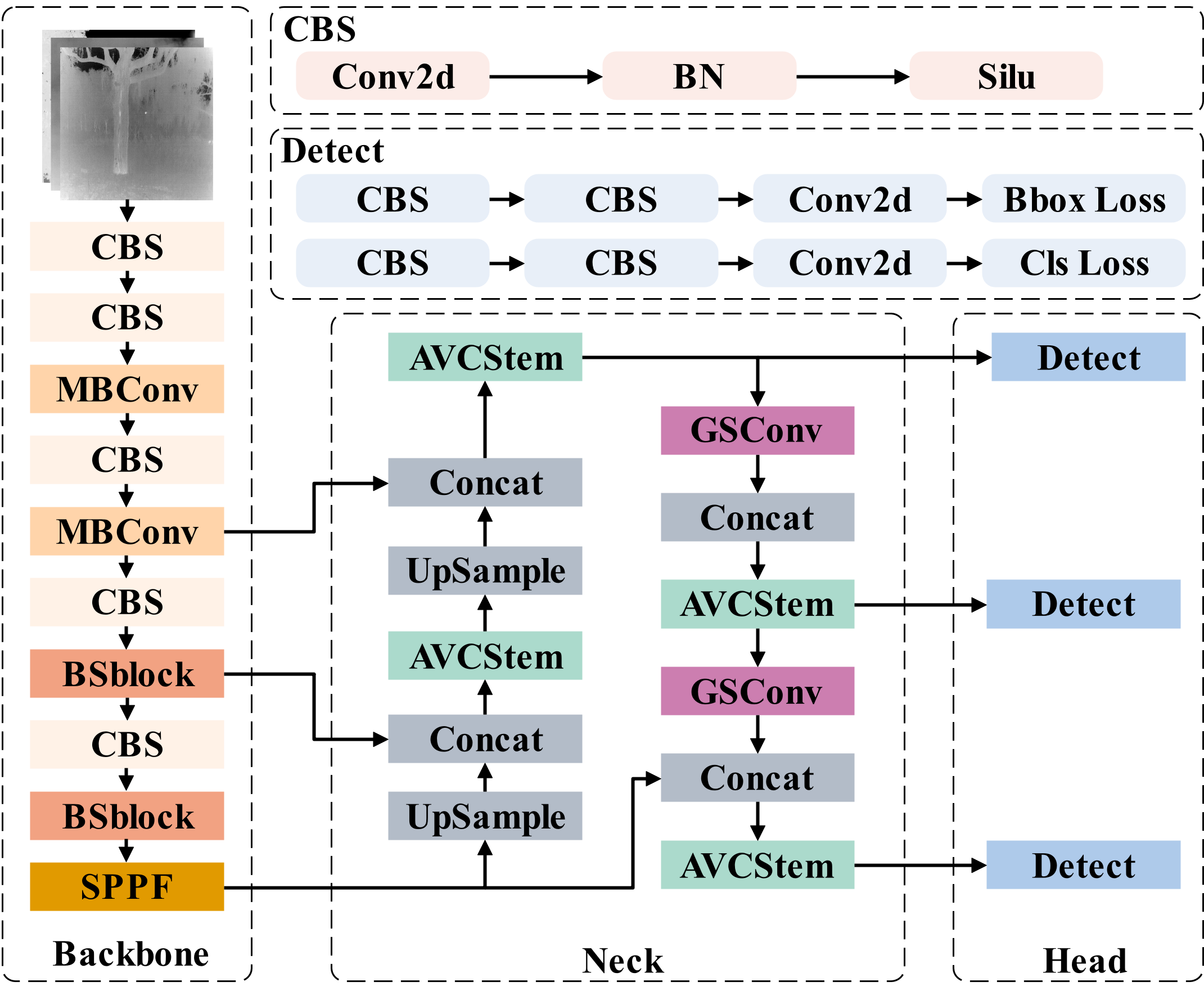}
    \caption{The overall framework of LE-IRSTD is shown in the figure, which includes the feature extraction backbone Faststem, the improved neck AVSFPN, and the detection head. The overall structure is an enhancement based on YOLOv8-n, achieving a balanced effect of precision and lightweight design.}
    \label{fig3:overall structure}
\end{figure*}

The Lightweight Efficient Infrared Small Target Detection (LE-IRSTD) framework builds upon the YOLOv8-n architecture, which have fewer parameters compared with the other YOLOv8 model and can maintain a certain degree of accuracy in IRSTD. The overall architecture of LE-IRSTD is shown in Fig. \ref{fig3:overall structure}. Specifically, the systematic architectural innovations in four critical dimension: feature extraction, multi-scale feature fusion, spatial-channel processing, and loss optimization. The innovation contributions can be decomposed as follows:

\begin{enumerate}
    \item \textbf{Adaptive feature extraction:} The backbone architecture integrates \textbf{Mobile Inverted Bottleneck Convolution (MBConv) block} with compound scaling strategy that dynamically optimizes depth, width, and input resolution through neural architecture search. This optimization achieves parameter reduction compared to baseline while maintaining high accuracy. The introduced \textbf{Bottleneck Structure (BS) block} performs channel-wise partial convolution through depth-wise separable operations, effectively suppressing redundant feature interference with fewer Flops than standard convolution.
    
    \item \textbf{Multi-dimensional Feature Fusion:} For cross-scale feature integration, we propose an \textbf{Attention-based Variable Convolution Stem (AVCStem) module} that processes features through parallel spatial-channel attention pathways. This is augmented by deformable convolution kernels with learnable offset parameters $(\Delta x, \Delta y) \in \mathbb{R}^2$ that adaptively adjust receptive fields according to target morphology. 

    \item \textbf{Information-Preserving Downsampling:} The \textbf{Global Shuffle Convolution (GSConv) module} addresses spatial information loss through hybrid depthwise-separable convolution with channel shuffle operations. 

\end{enumerate}

\subsection{Module-wise Improvements In LE-IRSTD}

\subsubsection{\textbf{MBConvblock}}
\begin{figure}
    \centering
    \includegraphics[width=0.6\linewidth]{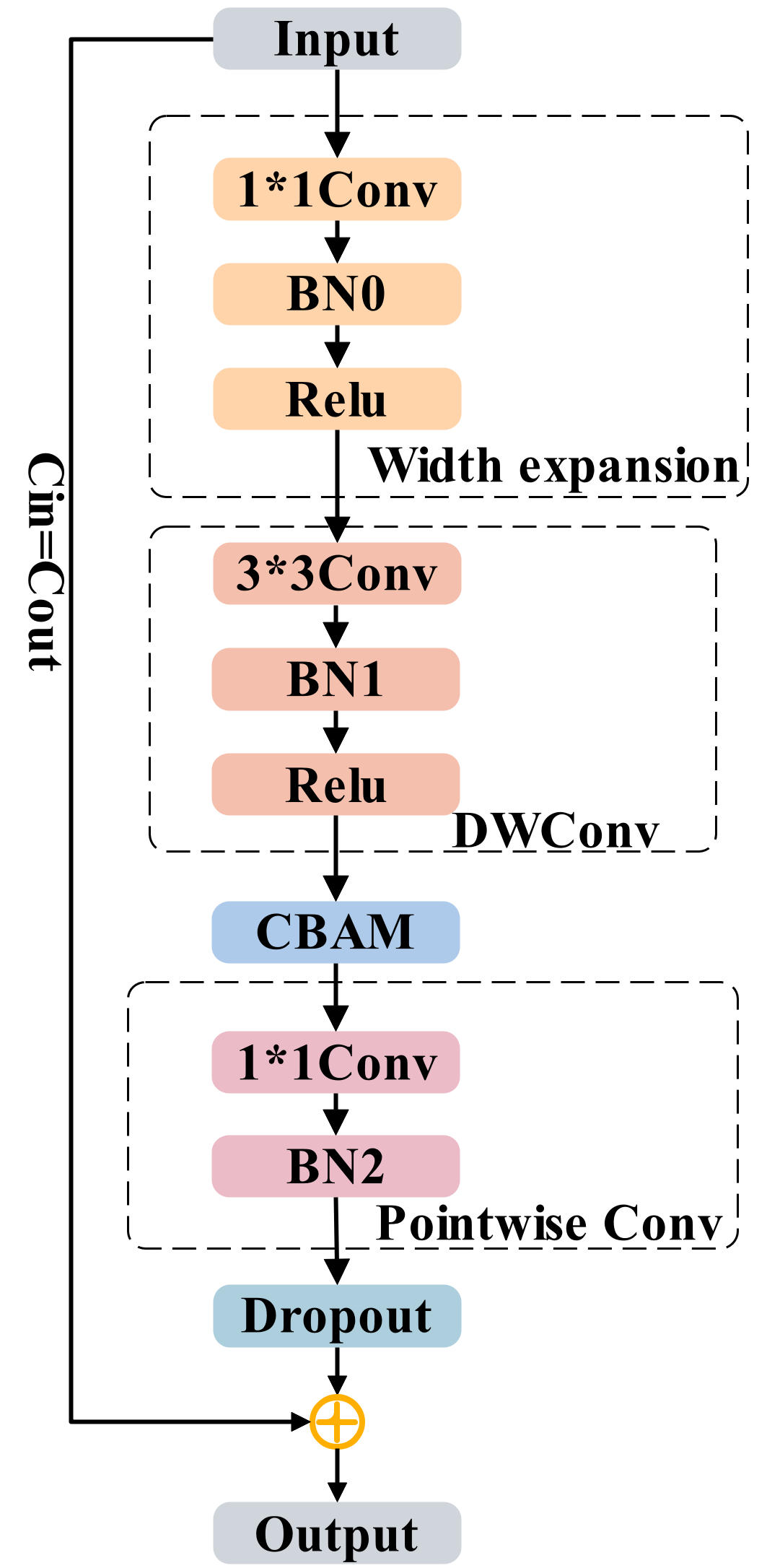}
    \caption{Structure of Mobile Inverted Bottleneck Convolution. Box 1 represents width expansion, Box 2 represents depthwise convolution, CBAM stands for Channel and Spatial Attention Mechanism, and when the input and output channels are equal, a residual connection is executed}
    \label{fig4:MBconvblock}
\end{figure}

To capture complex semantic information between small target and background, as well as overcome the challenges faced by methods such as increasing the number of convolutional layers or expanding the network width. We proposed Mobile Inverted Bottleneck Convolution block (MBConvblock), which is shown in Fig. \ref{fig4:MBconvblock}. MBConvblock utilizes different convolutions to change the width and resolution of the feature map. Firstly, A $ 1\times1 $ convolution is used to increase the dimensionality of input features $ X \in \mathbb{R}^{H \times W \times C} $ by changing the number of channels (width). In this paper, we set channels to 6 to expand the feature map channels, $X_{1}= Conv_{1\times1}X\in R^{H\times W\times 6C} $, where each channel specializes in extracting different features (e.g., edges, shapes, or texture details).
\begin{equation}
    \mathbf{X}_{1}=Conv_{1 \times 1}(\mathbf{X}) \in \mathbb{R}^{H \times W \times 6C}
\end{equation}


Secondly, depthwise convolutions modify the height and width parameters (resolution) of feature maps in the backbone stage compared to the C2f block in YOLOv8-n. A $3 \times 3$ depthwise convolution is then applied to adjust the height and width parameters of the feature map($X_{2}= DConv_{3\times3}X_{1}  \in R^{H{}' \times W{}'\times 6C} $), enabling adaptation to feature connections at different layers. The CBAM attention mechanism \cite{woo2018cbam} adaptively focuses on critical channel and spatial positions, improving detection precision by emphasizing important features. Finally, dropout is used to randomly discard some neurons, reducing interference from redundant features. 
\begin{equation}
\mathbf{X}_{out}= \left \{ _{\text{Drop}(\text{BN}(Conv_{1\times1}(\text{CBAM}(\mathbf{X}_{2} ))))+\mathbf{X}_{in} \;\;\;\;C_{in}=C_{out}}^{\text{Drop}(\text{BN}(Conv_{1\times1}(\text{CBAM}(\mathbf{X}_{2} )))) \;\;\;\;\;\;\;\;\;\;\;\;\;C_{in}\ne C_{out}}  \right \}
\end{equation}

The MBConvblock was designed to reduce the model's depth while aggregating scale information through various bottleneck modules, enabling channel-specific feature localization and deep semantic extraction. Overall, the model reaches saturation, the gradient vanishing problem arises, making it difficult for wider networks to capture high-level features and increasing the risk of model overfitting. Additionally, while higher input image resolution allows the model to capture more information, extremely high resolutions can lead to saturation in detection accuracy and excessive computational cost.


\subsubsection{\textbf{BSblock}}
While Flops reduction remains a common optimization target for efficient and real-time IRSTD, we observe through rigorous latency analysis that simply minimizing Flops often fails to achieve practical speed improvements \cite{chen2023run}. This paradox primarily stems from conventional architectures' tendency to stack excessive convolutional layers for depth enhancement, inadvertently introducing memory access bottlenecks. We proposed Bottleneck Structure block(BSblock), which leverages Partial Convolution (PConv) to strategically process spatial features from a channel subset while preserving information flow through intact channels, effectively mitigating redundant computational costs and memory access overhead.


\begin{figure}
    \centering
    \includegraphics[width=0.6\linewidth]{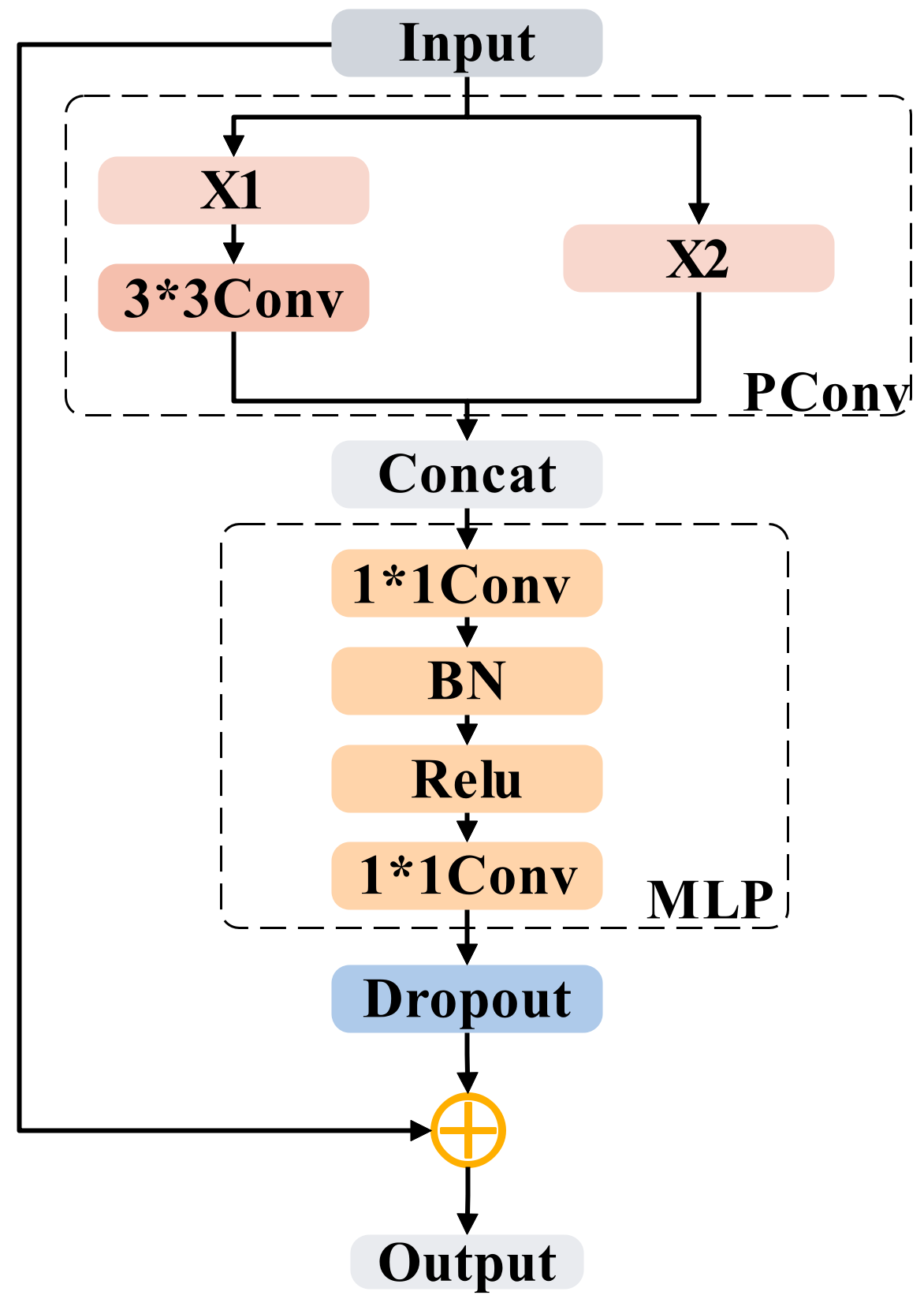}
    \caption{Overview of Bottleneck Structure block. PConv stands for Partial Convolution, MLP represents Multilayer Perceptron, and after performing dropout, it connects with the input features to achieve feature information aggregation.}
    \label{fig5:BSblock}
\end{figure}

As illustrated in Fig. \ref{fig5:BSblock}, BSblock architecture implements channel-wise separation through three optimized operations: First, input features undergo channel partitioning where a representative subset (r = 1/4) undergoes spatial processing via $3\times3$ convolutions. The processed features then combine with bypassed channels through channel-wise concatenation, forming the input for subsequent MLP transformations. To enhance robustness and reduce over-fitted, we implement stochastic regularization through dropout (p = 0.2) before establishing residual connections—a dual mechanism that simultaneously prevents gradient vanishing and improves generalization performance. Mathematically, this process can be formulated as:
\begin{equation}
    \mathbf{X}^{\prime}=\mathbf{X}+\text{Drop}(\text{MLP}(\text{Concat}(Conv(\mathbf{X}(\frac{c}{4},H,W)), \mathbf{X}(\frac{3c}{4},H,W))))
\end{equation}



The Partial Convolution operation achieves computational efficiency through channel selectivity:
\begin{equation}
\text{PConv}(\mathbf{X}) = \text{Concat}[Conv_{3 \times 3}(\mathbf{X}{:C_p }), \mathbf{X}_{C_p:}]
\end{equation}
Where $\mathbf{X}$ is input feature map, $C_p$ is the channels of $\lbrace 0 \xrightarrow{} C_p \rbrace$ of input feature map $\mathbf{X}$.

By maintaining contiguous memory access patterns through sequential channel processing, PConv reduces memory access costs to 25\% of standard convolution when using our recommended compression ratio $r=C_{p}/C=1/4$. Flops analysis reveals more dramatic savings:
\begin{equation}
\text{Flops}({\text{PConv}}) = r^2 \cdot C_p^2 = \frac{1}{16}\text{Flops}({\text{StdConv}}), (r=1/4)
\end{equation}
Where PConv is partial convolution operation, StdConv is standard convolution operation. $C_p$ is the channels of $0-C_p$ of input feature map.

This sophisticated design enables our backbone network to maintain baseline accuracy while reducing computational costs by compared to conventional architectures. Notably, the channel preservation strategy ensures no representational capacity loss, achieving an optimal balance between efficiency and performance.







\begin{figure}
    \centering
    \includegraphics[width=1\linewidth]{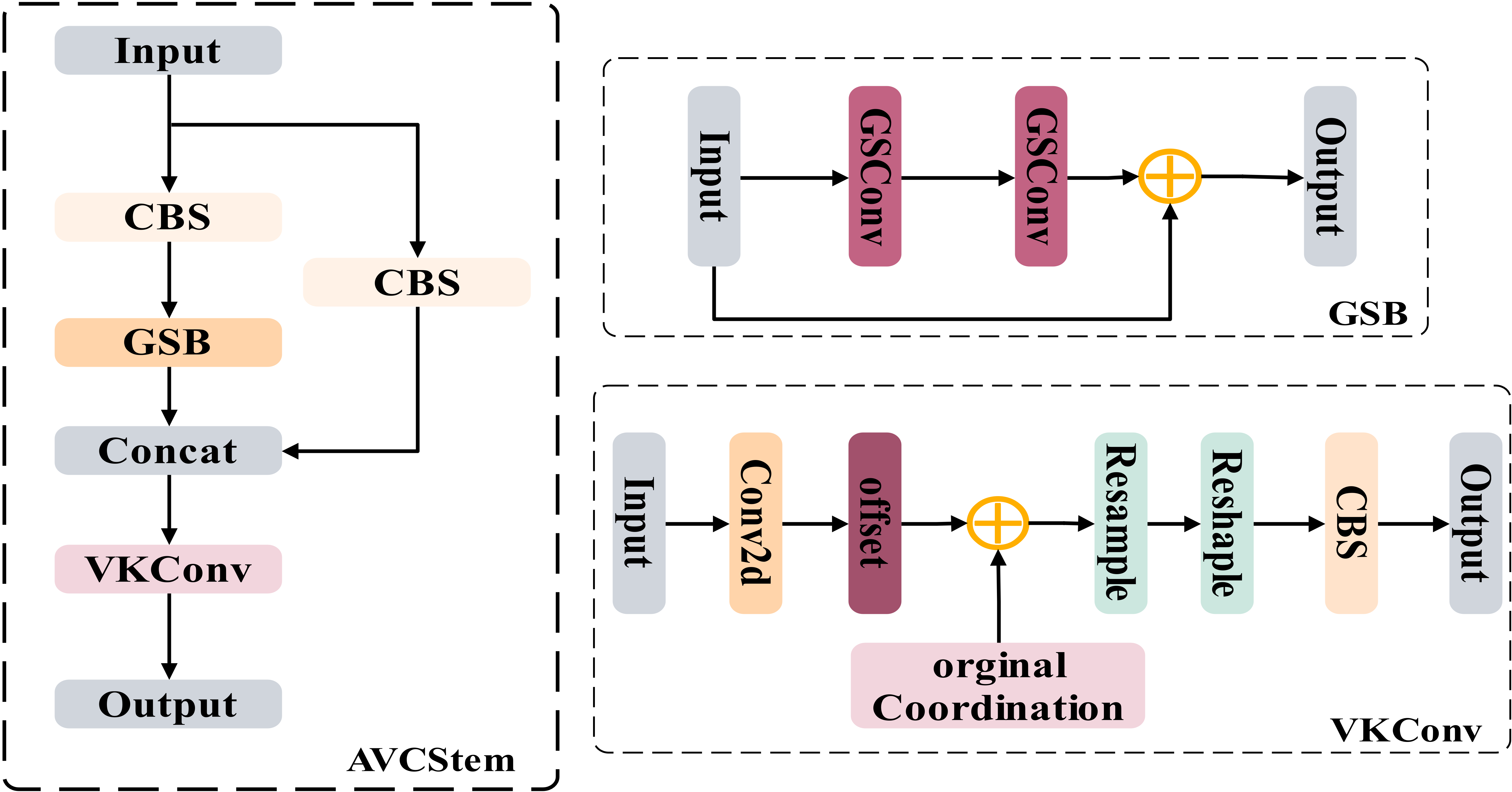}
    \caption{Overview of the Attention-based Variable Convolution Stem. In the parallel channels, CBS refers to the standard convolution operation, which includes batch normalization and the silu activation function. GSB connects with the input features after performing two GSConv operations. VKConv generates different sampling shapes by adding the base coordinates and offsets. Finally, the data and matrix shapes are adjusted through resample and reshape to obtain the final output.}
    \label{fig6:AVCStem}
\end{figure}

\subsubsection{\textbf{AVCStem}}

As depicted in Fig. \ref{fig6:AVCStem}, our enhanced YOLOv8-n architecture replaces conventional C2f modules with the Attention-based Variable Convolution Stem (AVCStem) module. This innovative component employs parallel convolutional branches to concurrently process features across divergent spatial-channel domains, achieving faster cross-dimensional fusion than baseline implementations. The design integrates three critical innovations:
\begin{enumerate}
    \item \textbf{Global Synapse Branch:} Facilitates inter-channel information exchange through dynamic attention weighting, mathematically expressed as:
    \begin{equation}
    \mathbf{W}g = \sigma(\mathcal{F}_{1 \times 1}(\mathbf{X}) \otimes \mathcal{F}_{3 \times 3}(\mathbf{X}))
    \end{equation}
    Where $\sigma$ denotes sigmoid activation and $\otimes$ represents Hadamard product.
    \item \textbf{Multi-branch Feature Extraction:} Implements complementary spatial processing through parallel $3\times3$ depthwise and $1\times1$ pointwise convolutions.
    \item \textbf{Adaptive Kernel Adjustment:} Incorporates Variable-shape Kernel Convolution (VKConv) \cite{zhang2024ldconv} during feature concatenation, enabling real-time kernel geometry adaptation through:
\end{enumerate}

\textbf{Phase 1: Base Coordinate Initialization:}Establishes fundamental sampling positions:
\begin{equation}
P_n = \mathcal{G}(K_s, S) \in \mathbb{R}^{5 \times 2}
\end{equation}
where $K_s$ denotes kernel size ($5 \times 5$ default) and $S$ indicates stride parameters.

\textbf{Phase 2: Adaptive Offset Prediction:}Learns position-specific deformations using input-dependent offsets:
\begin{equation}
\Delta P = \mathcal{F}_{adapt}(\mathbf{X}) \cdot \alpha \quad (\alpha=0.1)
\end{equation}
where learnable scaling factor $\alpha$ initialized at 0.1.

\textbf{Phase 3: Dynamic Resampling:}Executes deformable feature sampling through coordinate adjustment:
\begin{equation}
\mathbf{X}_{out} = \sum \lbrace p\in | P_0+\Delta P \cdot \mathbf{X}_{in}(p) \rbrace \cdot \mathbf{W}_p
\end{equation}

\begin{figure}
    \centering
    \includegraphics[width=1\linewidth]{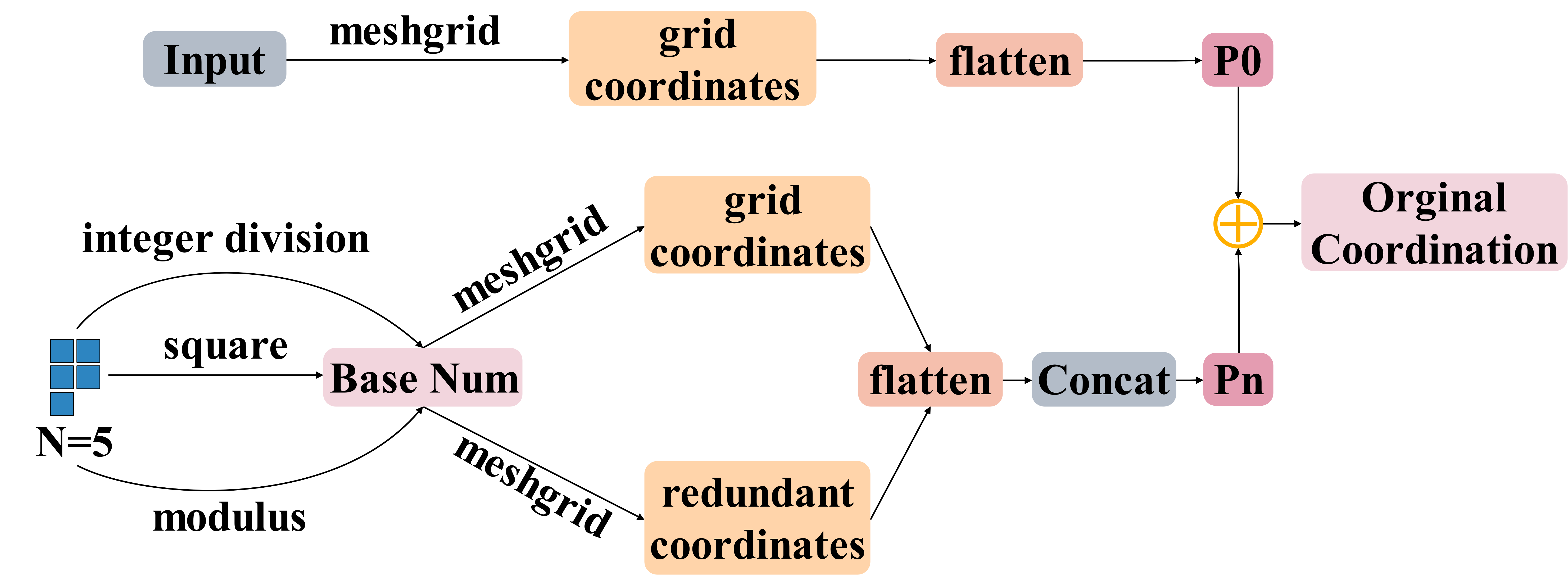}
    \caption{The orginal coordinate generation algorithm is illustrated. The original convolution kernel parameters undergo a square root process to derive the base parameters. The original parameters are computed using integer division and modulus operations. Next, a meshgrid is used to generate grid coordinates, which are then flattened into a one-dimensional shape to create the combined shape $Pn$. Finally, $P0$ is added to $Pn$ to obtain the final original sampling coordinates.}
    \label{fig7:VKConv}
\end{figure}

As visualized in Fig. \ref{fig7:VKConv}, this process overcomes fixed geometric constraints of conventional convolution through shape adaptivity, parametric efficiency, and computational Economy. Overall, the synergistic integration of AVCStem and VKConv achieves better metrics improvement on IRSTD while maintaining satisfied inference speed, demonstrating superior of "make both ends meet" with efficiency-accuracy.

\begin{figure}
    \centering
    \includegraphics[width=1\linewidth]{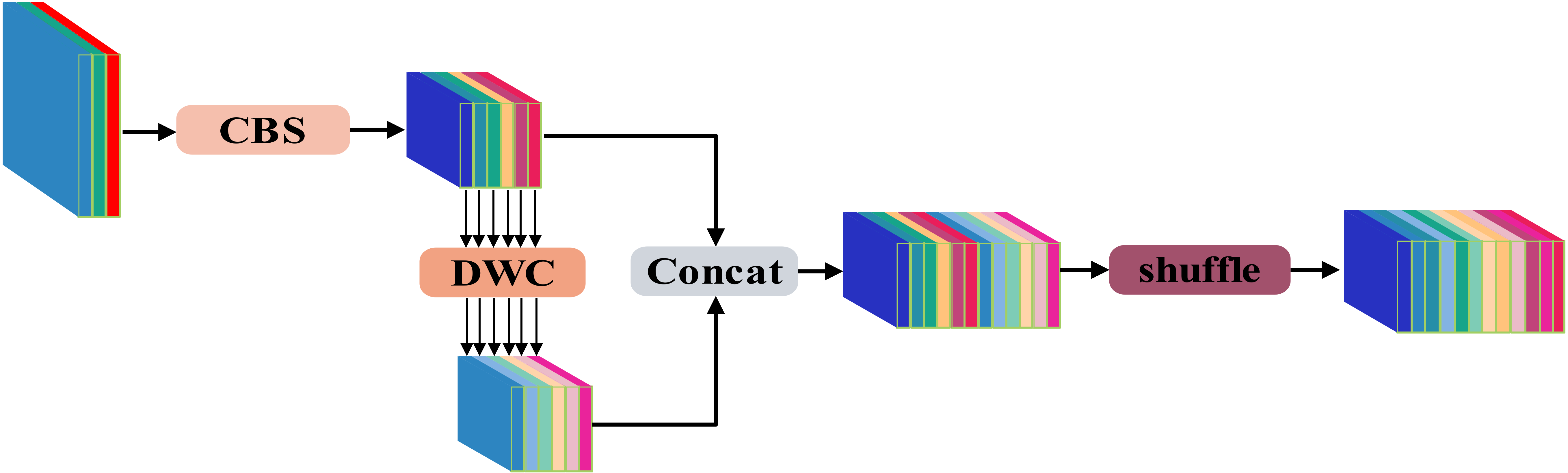}
    \caption{The structure of GSConv: the input first undergoes "Convolution, Batch normalization, and Silu activation" (CBS) for downsampling, followed by a depthwise convolution(DWC). The results of these two convolutions are then concatenated, and a shuffle operation is performed at the end}
    \label{fig8:GSConv}
\end{figure}

\subsubsection{\textbf{GSConv}}

Traditional backbone networks inevitably cause progressive semantic attenuation during channel expansion and spatial downsampling—particularly detrimental for IRSTD where low-contrast features require persistent propagation \cite{li2022dense}. This phenomenon stems from the architectural dilemma: dense convolution maintains implicit inter-channel connections at computational redundancy, while sparse convolution sacrifices feature interdependence for efficiency, the processing of GSConv are as follows:

\textbf{Phase 1: Feature Conditioning:}
\begin{equation}
\mathbf{F}_c = \text{SiLU}(\text{BN}(\mathcal{C}_{3 \times 3}(\mathbf{X}_{in}))) \in \mathbb{R}^{H \times W \times C}
\end{equation}
where $\mathcal{C}_{3 \times 3}$ denotes standard convolution with kernel size 3.

\textbf{Phase 2: Efficient Spatial Encoding:}
\begin{equation}
\mathbf{F}_d = \text{DWConv}_{3 \times 3}(\mathbf{F}_c) \in \mathbb{R}^{\frac{H}{2} \times \frac{W}{2} \times C}
\end{equation}
where DWConv applies single-channel kernels per input channel, preserving channel dimensionality.

\textbf{Phase 3: Feature Information Fusion:}
\begin{equation}
\mathbf{F}_{out} = \text{ChannelShuffle}(\text{Concat}[\mathbf{F}_c, \mathbf{F}_d]) \in \mathbb{R}^{\frac{H}{2} \times \frac{W}{2} \times 2C}
\end{equation}

The channel shuffle operation strategically interleaves features from both pathways, ensuring adjacent channel positions contain complementary information—crucial for maintaining gradient flow. Compared to standard $3\times3$ convolution, GSConv achieves Flops reduction while increasing feature diversity.

\subsection{Loss function}
Our overall loss function integrates classification-aware and geometry-aware components, the classification-aware loss function is Binary Cross-Entropy (BCE) loss, and the geometry-aware components complete IoU(CIoU) and Dynamic Focal Loss(DFL):

\textbf{Classification loss}:
BCE loss evaluates target presence probability through dual-case logarithmic penalties, ensuring non-zero gradients for both positive/negative samples:
\begin{equation}
   \mathcal{L}_{BCE}=-(ylog(p(x)+(1-y)log(1-p(x))
\end{equation}

\textbf{Regression Loss}: Combines Complete IoU (CIoU) and Dynamic Focal Loss (DFL) for enhanced localization:

\begin{equation}
\mathcal{L}_{\text{CIoU}} = 1 - \text{IoU} + \frac{d^2}{c^2} + \alpha v
\end{equation}
\begin{equation}
\alpha=\frac{v}{(1-IOU)+v} 
\end{equation}
\begin{equation}
v=\frac{4\cdot (\arctan \frac{W_{g} }{H_{g}}-\arctan \frac{W_{p} }{H_{p}})^{2} }{\pi ^{2} } 
\end{equation}
where d is the distance between the centers of the predicted box and the ground truth box, and c is the diagonal distance of the minimum enclosing rectangle of the intersection area between the predicted box and the ground truth box. The formula for $\alpha$  is as follows, where v is a correction factor used to further adjust the loss function. $(W_{g}, H_{g})$ and $(W_{p}, H_{p})$ represent the width and height of the ground truth box and the predicted box, respectively.

\begin{equation}
\mathcal{L}_{DFL} =-\alpha (1-p)^{\gamma } \cdot log(p)
\end{equation}
where p is the predicted probability, $\alpha$ is the balance factor used to adjust the weight of each class, and $\gamma$ is the focal factor used to enhance the focus on difficult samples.

The overall loss function is:
\begin{equation}
\mathcal{L}_{all} = \lambda_{1}\mathcal{L}_{BCE} + \lambda_{2}\mathcal{L}_{CIoU} + \lambda_{3}\mathcal{L}_{DFL} 
\end{equation}
where $\lambda_{i}(i=1,2,3)$ is weights parameter.
\AtBeginEnvironment{table}{\footnotesize} 

\section{Experiments and Analysis} \label{sec:experiment}
To comprehensively evaluate LE-IRSTD, we conducted a rigorous protocol across three open-source infrared datasets: NUAA-SIRST \cite{dai2021asymmetric}, IRSTD-1K \cite{zhang2022isnet}, and NUDT-SIRST \cite{li2022dense}. Our experimental design systematically address three critical research questions through comparative analysis with state-of-the-art (SOTA) methods and detailed component-wise ablation studies. The investigation framework is structured as follows:

\begin{enumerate}

\item{\textbf{Q1: Comparative Analysis With State-of-the-Art}}: We will compare LE-IRSTD with other SOTA methods, focusing on the computational efficiency such as complexity(M), Flops(G), detection accuracy such as Precision, Recall, F1-score, and mAP@50.

\item{\textbf{Q2: Lightweight Implementation Of The Model}}:We compared each module in LE-IRSTD with its baseline such as parameter and Flops reduction to verify the lightweight effect of each module.

\item{\textbf{Q3: Improved Model Performance}}: We employ Grad-CAM \cite{selvaraju2017grad} visualizations and quantitative ablation study to each block contribution to final detection performance, and analyze feature evolution patterns across network stages. Therefore, it validate information retention capabilities during downsampling.

\end{enumerate}

\subsection{Experimental Settings} \label{subsec:setting}

\subsubsection{\textbf{Dataset}}

\begin{figure}
    \centering
    \includegraphics[width=1\linewidth]{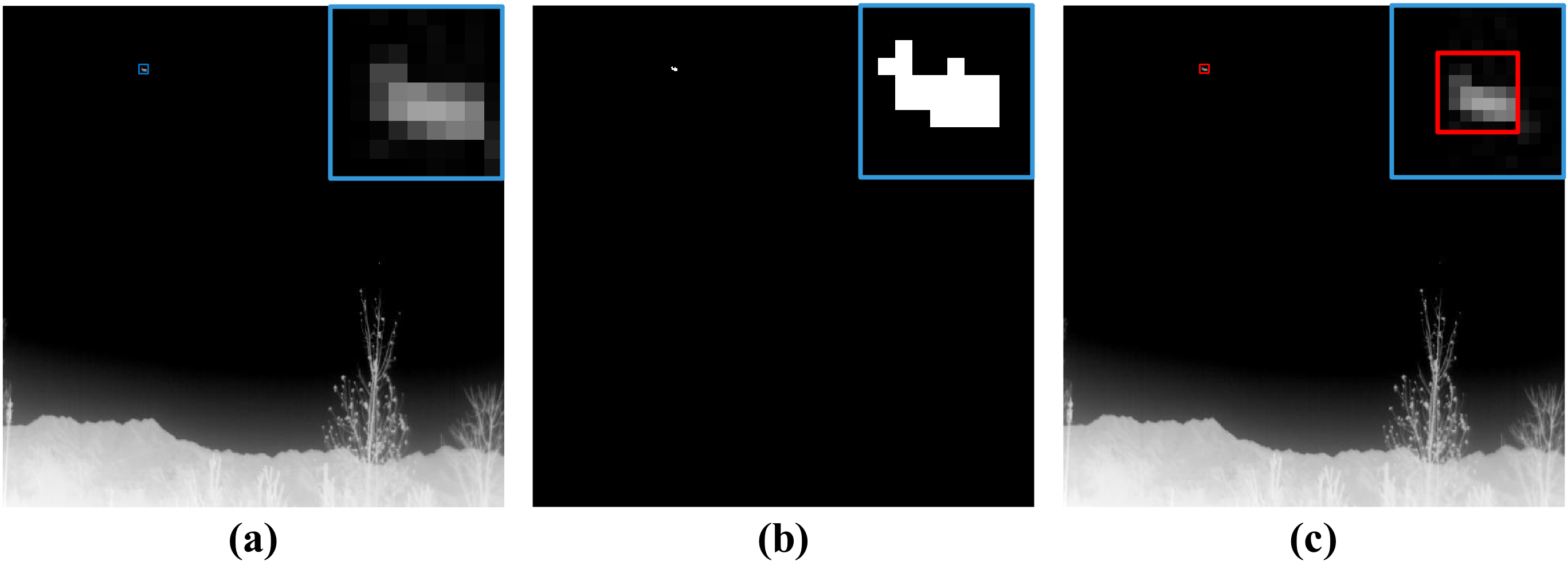}
    \caption{Different annotation forms for the current infrared small target public dataset. (a) Infrared small target. (b) Semantic label. (c) Bounding box label.
}
    \label{fig:6}
\end{figure}

We utilized bounding box annotations on three public datasets (YOLO format). IRSTD-1K \cite{zhang2022isnet}: A dataset focused on infrared small target detection, containing 1000 images with precise pixel-level annotations. NUAA-SIRST \cite{dai2021asymmetric}: A high-quality infrared small target detection dataset containing 427 real images accompanied by corresponding labels. NUDT-SIRST \cite{li2022dense}: A dataset specifically designed for infrared small target detection, containing 1327 training and validation examples. It is designed to provide rich resources for the field of infrared small target detection, as illustrated in Fig. \ref{fig:6}, which depicts the detection framework and segmentation mask approach for small and medium targets in infrared images. Each dataset was divided into a training set (60\%), validation set (20\%), and test set (20\%).

\subsubsection{\textbf{Evaluation Metrics}}
To compare the proposed method with the state-of-the-art (SOTA) methods, we employ commonly used evaluation metrics including precision, recall, and F1. Each metric is defined as follows:
\begin{equation}
    \text{Precision} = \frac{TP}{TP + FP}
\end{equation}

recall measure the ability to discover all positive cases, calculated as the ratio of TP to the sum of TP and false negatives (FN).
\begin{equation}
    \text{Recall} = \frac{TP}{TP + FN}
\end{equation}

F1 is a harmonic mean of precision and recall, providing a balanced measure of the model’s performance, which is computed as:
\begin{equation}
    \text{F1} = \frac{2 \times (\text{Precision} * \text{Recall})}{\text{Precision} + \text{Recall}} 
\end{equation}

$TP$(True Positive) : The actual correct is classified as the correct number of samples. $FP$(False Positive) : Actual errors are classified as the correct number of samples. $TN$(True Negative) Actual errors are divided into incorrect sample numbers. $FN$(False Negative) : The number of samples that are actually correctly classified as errors.

The average precision (AP) is the average of the accuracy at different recall points, represented on the PR curve as the area below the PR curve. The larger the value of AP, the higher the average accuracy of the model.
\begin{equation}
    AP = \int_{0}^{1} \text{Precision}(\text{Recall}) \, d(\text{Recall})
\end{equation}

mAP@0.5 When bounding box IoU is set to 0.5, calculate the AP of all images in each category, and then average all categories. In the experimental section, mAP is used to represent mAP@50.
\begin{equation}
    \text{mAP} = \frac{1}{N} \sum_{i=1}^{N} AP_i
\end{equation}
where $AP_i$ represents the $AP$ value of the i-th category, and $N$ represents the number of categories in training dataset.


Mean Normalized Contrast Average Precision (mNoCoAP) \cite{dai2023one} is a novel evaluation metric for infrared small target detection, inspired by the widely used mAP in object detection tasks. It offers a unified, fair, and robust evaluation standard across different detection paradigms, such as bounding box regression, semantic segmentation, and background suppression. The core principle of mNoCoAP lies in its definition of true positives: a prediction is considered a true positive if the NoCo (Normalized Contrast) value of the predicted target region exceeds a certain threshold. The parameter $\delta$ represents the NoCo threshold that controls the required precision of centroid localization. In contrast, the traditional mAP defines true positives based on the Intersection over Union (IoU) between the predicted and ground truth boxes, typically requiring IoU > 0.5.

\begin{equation}
    \text{NoCo} = \frac{\mu_T - \mu_B}{\sigma_B}
    \label{eq:noco}
\end{equation}

$\mu_T$: the mean gray value of pixels within the predicted target region, $\mu_B$: the mean gray value of the background region (typically an annular area surrounding the target), $\sigma_B$: the standard deviation of the background region.
\begin{equation}
    \text{mNoCoAP} = \frac{1}{9} \sum_{\delta=0.1}^{0.9} \text{AP}_\delta
    \label{eq:mNoCoAP}
\end{equation}

Moreover, The Parameter and Flops computation are used to describe the complexity of the model.

\subsubsection{\textbf{Experimental Details}}

All experiments achieved convergence after being trained for 300 epochs. The models were trained with a fixed input image resolution of $640 \times 640$ pixels using the Adam optimizer \cite{kingma2014adam}. The training configuration included batch size of 16, momentum of 0.937, and weight decay of 0.0005. To stabilize training, a warm-up phase of 3 epochs was set, with the optimizer momentum at 0.8. After the warm-up, the learning rate decayed using a cosine annealing function, with an initial learning rate of 0.001 and a final learning rate of 0.5. The experimental hardware setup consisted of 12 vCPUs on Intel Xeon Platinum 8352V processors running at 2.10 GHz, equipped with an NVIDIA GeForce RTX 4090 GPU (24GB VRAM) and running Ubuntu 18.04. The software stack included PyTorch 1.8.1, Python 3.8, and CUDA 11.1.


\subsection{Comparison with State-of-the-Arts} \label{subsec:sota}

Compared to traditional methods such as NewTopHat \cite{bai2010analysis}, FKRW \cite{qin2019infrared}, TLLCM \cite{han2019local}, MPCM \cite{wei2016multiscale}, IPI \cite{gao2013infrared}, and RIPT \cite{dai2017reweighted}, deep learning approaches demonstrate significantly superior performance. There are primarily two types of deep learning-based methods for object detection: segmentation-based methods, including AGPCNet \cite{zhang2021agpcnet}, ALCNet \cite{dai2021attentional}, and SCTransNet \cite{yuan2024sctransnet}. These methods achieve object detection by separating the target foreground from the distracting background in images. However, current semantic segmentation-based detection methods are unable to accurately segment and detect small targets with clear contours. The second type is detection-based methods, such as R-CNN \cite{girshick2014rich}, Faster R-CNN \cite{ren2016faster}, RetinaNet \cite{ross2017focal}, AMFLW-YOLO \cite{peng2023amflw}, DET-YOLO \cite{chen2024det} and EFLNet \cite{yang2024eflnet}, OSCAR \cite{dai2023one}, which enhance the probability of detecting small targets by precisely locating objects within detection boxes. To ensure a fair comparison, each model was retrained using the three public datasets for 300 epochs, with all other parameters maintained at their default values.

\subsubsection{\textbf{Quantitative Results and Analysis}}


Tab. \ref{tab:sota} presents a quantitative comparison of LE-IRSTD with current state-of-the-art (SOTA) approaches. The LE-IRSTD method demonstrates superior performance across all evaluation metrics in the IRSTD-1k \cite{zhang2022isnet}, NUAA-SIRST \cite{dai2021asymmetric}, and NUDT-SIRST \cite{li2022dense} datasets.
In the IRSTD, most deep learning-based methods treat this task as a pixel-level segmentation problem, requiring accurate pixel-level segmentation outputs. Any deficiency in this process can lead to false positives or missed detections, negatively impacting overall detection performance at the object level.
Additionally, these methods \cite{chen2024det, peng2023amflw} has limited focus on addressing the feature-preserving with infrared small objects and sensitivity issues related to bounding boxes. Consequently, performance in relatively low precision, recall, and F1 scores.
In contrast, our LE-IRSTD employs a bounding-box regression strategy and efficient feature preserving that has achieved significant improvements enables precise localization and detection performace in IRSTD compared to other methods.

\begin{table*}[ht]
  \renewcommand\arraystretch{1.2}
  \centering
  \caption{Comparison with Other State-of-the-art methods on three datasets. The $\uparrow$ indicates that the higher the indicator, the better. We display the best result in the \textcolor{red}{red} color and the second-best result in the \textcolor{blue}{blue} color, the third-best result in \textcolor{green}{green} color. mAP represents mAP@0.5 with a threshold of 0.5, while ncap represents mNoCoAP}
  \label{tab:sota}
  \setlength{\tabcolsep}{3pt}
  
  \begin{tabular}{c|c|c|ccccc|ccccc|ccccc}
    \multirow{2}{*}{Method} & \multirow{2}{*}{Para$\downarrow$} &\multirow{2}{*}{Flops$\downarrow$}  & \multicolumn{5}{c|}{IRSTD-1K(Tr=60\%)}  & \multicolumn{5}{c|}{NUAA-SIRST(Tr=60\%)}   & \multicolumn{5}{c}{NUDT-SIRST(Tr=60\%)} \\
       & & &  P $\uparrow$ & R$\uparrow$ & F1 $\uparrow$ & mAP $\uparrow$ & ncap $\uparrow$ & P $\uparrow$ &  R $\uparrow$ & F1 $\uparrow$ & mAP $\uparrow$ & ncap $\uparrow$ & P $\uparrow$ &  R $\uparrow$ & F1 $\uparrow$ & mAP $\uparrow$ & ncap $\uparrow$\\
  \hline
  \multicolumn{13}{l}{\textit{Background Suppression}}  \\
  \hline
  FKRW \cite{qin2019infrared} &- &- & 43.6 & 57.3 & 49.5 & 38.0 & 27.8 & 50.7 & 63.7 & 56.5 & 78.8 & 30.4 & 59.5 & 42.1 & 49.3  & 60.8 & 38.7\\ 
  NewTopHat \cite{bai2010analysis} &- &- & 49.2 & 50.2& 49.7 & 49.2 &30.5 & 71.2 & 43.9 & 54.3 & 80.3 & 31.4 & 52.7 & 60.2 & 56.2 & 59.9 & 42.1 \\
  \hline  
  \multicolumn{13}{l}{\textit{HVS-based Methods}} &  \\  
  \hline
  MPCM \cite{wei2016multiscale} &- &- & 43.5 & 53.2 & 47.93  & 45.0 &36.9 & 38.6  & 36.8 & 37.7  & 40.1 &30.1 & 72.6 & 89.1 & 80.0  & 80.7 & 47.3\\
  TLLCM \cite{han2019local} &- &- & 58.3 & 77.6 & 66.6  & 75.4 &50.4 & 82.5 & 72.2 & 77.0  & 62.1 & 54.2 & 43.8 & 73.1 & 54.8  & 50.2 & 41.6\\
  GSWLCM \cite{qiu2022global} &- &- & 54.2 & 60.3 & 57.1  & 70.2 & 51.8& 72.6 & 62.7 & 67.3 & 72.2 & 58.1& 12.7 & 20.3 & 15.7  & 40.7 & 29.2\\
  \hline
  \multicolumn{13}{l}{\textit{Low-rank and Sparse Decomposition}}  \\
  \hline
  IPI \cite{gao2013infrared} &- &- & 70.4 & 48.2 & 57.3  & 78.9 & 37.8 & 83.4 & 82.3 & 82.9  & 70.3 & 58.7 & 79.6 & 83.5 & 81.5  & 70.8 & 60.9 \\
RIPT \cite{dai2017reweighted}  &- &- & 68.6 & 80.4 & 74.0  & 69.8 & 30.8 & 79.5 & 63.2 & 70.4  & 83.2 & 43.8 & 78.0 & 60.3 & 68.0  & 79.5 & 78.3 \\
PSTNN \cite{zhang2019infrared} &- &- & 70.4 & 80.6 & 75.1  & 66.4 & 47.9 & 77.2  & 92.2  & 84.1 & 67.3 & 63.9 & 60.2  & 24.2 & 34.5 & 74.2 & 73.7 \\
\hline
\multicolumn{13}{l}{\textit{Deep Learning Methods}}  \\
\hline
R-CNN \cite{girshick2014rich} &38.5M &67.8G & 76.6 & 60.7 & 67.7 & 65.4 & 53.8 & 76.8 & 76.2 & 76.5 & 78.8 & 78.2 & 73.3 & 75.4 & 74.3 & 70.5 & 73.5 \\
Faster R-CNN \cite{ren2016faster} & 33.0M &75.9G & 82.4 & 84.8 & 83.6 & 83.3 & 70.0 & 85.3 & 80.4 & 82.8 & 89.3 & 84.3 & 88.9 & 89.3 & 89.1 & 83.2 & 82.2 \\
AGPCNet\cite{zhang2021agpcnet} & 12.4M &270.4G & 83.3 & 75.7 & 79.3 & 73.2 & 72.2 & 82.4 & 84.7  & 83.5 & 85.3 & 82.2 & 74.3 & 75.6 & 75.0 & 72.2 & 68.9 \\
ALCNet \cite{dai2021attentional} &20.2M &34.3G & 86.5  & 80.2 & 83.3 & 84.3 & 87.3 & 85.2  & 89.2  & 87.1  & 84.8 & 80.5 & 76.3 & 87.4 & 81.5  & 90.4 & 85.4 \\
RetinaNet \cite{ross2017focal} & 36.3M &75.9G & 83.9 & 83.9 & 83.9 & 89.5 & 81.0 & \textcolor{green}{89.3} & 78.9 & 83.8 & 84.3 & 80.3 & 83.1 & 85.8 & 84.4 & 85.2 & 81.2 \\
SCTransNet \cite{yuan2024sctransnet} &11.2M &20.2G & \textcolor{red}{91.7}  & 80.9 & 85.9  & 92.2 & 81.2 & 82.9  & 84.8  & 83.9  & 85.7 & 83.2 & \textcolor{green}{90.2}  & 88.5 & \textcolor{green}{89.3} & \textcolor{blue}{93.3} & 70.6 \\
AMFLW-YOLO \cite{peng2023amflw} &2.0M &4.6G & 87.5 & \textcolor{blue}{90.7} & \textcolor{blue}{89.0} & 93.3 & 83.8 & 88.8 & \textcolor{blue}{89.4} & \textcolor{green}{89.1}  & 83.9 & 83.3 & 89.4 & 84.9 & 87.1 & 90.2 & 84.4 \\
DET-YOLO \cite{chen2024det} & 41.6M &24.9G & 84.5 & 85.3 & 84.9 & \textcolor{green}{94.9} & \textcolor{blue}{85.6} & 86.3 & 88.3 & 87.3  & 84.8 & \textcolor{blue}{85.3} & 84.2 & \textcolor{red}{91.8} & 87.8  & 85.0 & 86.2 \\
EFLNet \cite{yang2024eflnet} & 38.3M & 103.6G & 87.0 & 81.7 & 84.3 & 93.9 & 84.3 & 88.3 & 85.3 & 86.8 & \textcolor{green}{92.2} & 85.3 & 86.2 & \textcolor{blue}{91.5} & 88.8  & \textcolor{green}{91.4} & \textcolor{blue}{88.7} \\
OSCAR \cite{dai2023one} & 42.2M & 68.3G & \textcolor{green}{88.5} & \textcolor{green}{89.3} & \textcolor{green}{88.9} & \textcolor{blue}{95.0} & \textcolor{green}{85.0} & \textcolor{blue}{90.9} & \textcolor{green}{89.1} & \textcolor{blue}{90.0}  & \textcolor{blue}{93.4} & \textcolor{blue}{86.1} & \textcolor{blue}{91.3} & 90.3 & \textcolor{blue}{90.8} & 90.8 & \textcolor{green}{88.6} \\
\hline
\rowcolor[rgb]{0.9,0.9,0.9}$\star$ \textbf{LE-IRSTD (Ours)}  &2.6M &7.4G & \textcolor{blue}{89.8}  & \textcolor{red}{91.2} & \textcolor{red}{90.5} & \textcolor{red}{95.4} & \textcolor{red}{87.2} &  \textcolor{red}{92.2}  & \textcolor{red}{90.4}  & \textcolor{red}{91.3}  & \textcolor{red}{95.3} & \textcolor{red}{86.5} & \textcolor{red}{93.2} & \textcolor{green}{90.8} & \textcolor{red}{92.0} & \textcolor{red}{94.1} & \textcolor{red}{89.2} \\
\end{tabular}
\end{table*}

\subsubsection{\textbf{Qualitative Visualization and Analysis}}



\begin{figure*}
    \centering
    \includegraphics[width=1\linewidth]{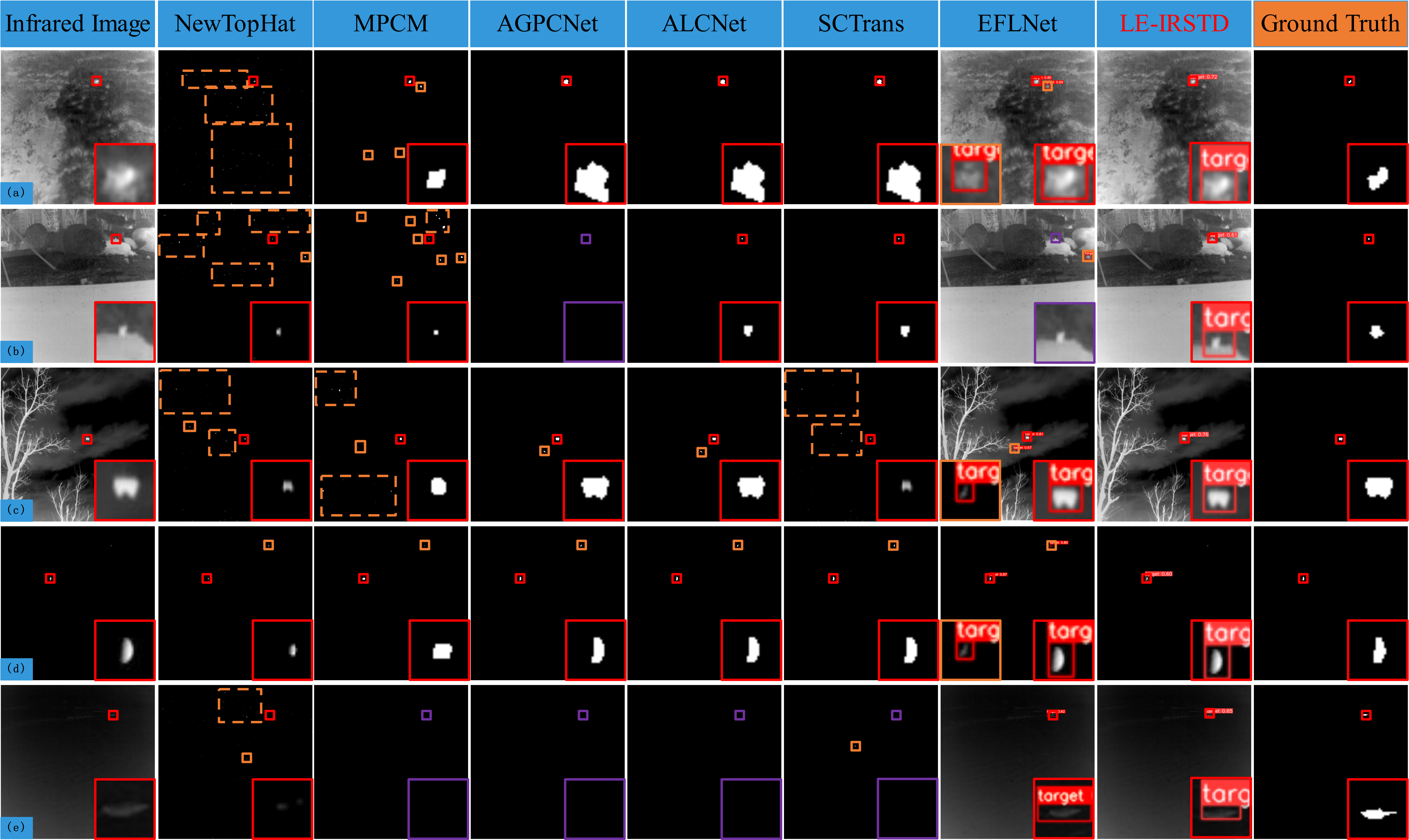}
    \caption{Partial visual results obtained from different methods on the IRSTD-1k dataset are shown. In the figure, the red, orange, and purple circles represent correctly detected targets, falsely detected targets, and missed targets, respectively.}
    \label{fig10:qualitative}
\end{figure*}

To rigorously evaluate detection robustness, we curated challenging scenarios from all three datasets exhibiting: \textbf{Cluttered backgrounds} (high-intensity interference patterns), \textbf{Low signal-to-noise ratios} (target signatures < 3dB), and \textbf{Prevalent distractors} (structural false alarm sources). As visualized in Fig. \ref{fig10:qualitative}, detection outcomes are annotated as: True positives (\textcolor{red}{red} boxes): Correctly detection targets, False positives (\textcolor{orange}{orange} boxes): Miss detection targets, and False negatives (\textcolor{purple}{purple} boxes): Missed detection targets.

Traditional methods produce a significant number of false detections and missed targets but also fail to focus on the target areas specifically, demostrate fundamental limitations. 
While deep learning methods shows substantial improvement over traditional methods, there is still room for enhancement, especially in complex background environments (as seen in the SCTransnet method's performance in Fig. \ref{fig10:qualitative} (c), (e), which exhibit false alarm and missed detections). Methods based on semantic segmentation still have suboptimal shape fidelity (Dice < 0.5) due to the boundary ambiguity in low-resolution targets (as illustrated in  Fig. \ref{fig10:qualitative} (a), where the shapes obtained from semantic segmentation differ significantly from the Ground Truth). In contrast, our LE-IRSTD method, which is based on object detection bounding boxes, is better suited for locating small targets within the original image, thereby enhancing detection accuracy.

\subsection{Ablation Study} \label{subsec:ablation}
\subsubsection{\textbf{Ablation of Different Model-wise Components}} \label{subsec:lamablation} 
We selected the IRSTD-1k \cite{zhang2022isnet} dataset as our ablation studies dataset due to its composition of real-world images and its substantial size. In contrast, the NUAA-SIRST dataset \cite{dai2021asymmetric} contains a limited number of images, while the NUDT-SIRST dataset \cite{li2022dense} consists of simulation-generated images.


See Tab. \ref{tab:LE-IRSTD-ablation} for specific results:
(1) Each module enhanced detection performance relative to the baseline while reducing computational overhead. These results demonstrate the feasibility of achieving lightweight implementation and improved detection accuracy, as proposed in this study.
(2) After incorporating the MBConvblock: (b) module appear increase in model parameter and Flops is attributed to the MBConv block in LE-IRSTD expanding the width to six times its original value, thereby extracting more features. Concurrently, compared to the baseline mAP of 86.4\%, the mAP increased to 89.4\% after introducing the MBConv block, which enhances the model’s detection accuracy.
(3) With the addition of the remaining modules, the detection accuracy ultimately reached 95.4\%. Overall, the model’s performance improved while the parameter and Flops decreased.


\begin{table*}
  \setlength{\abovecaptionskip}{0cm}  
  \renewcommand\arraystretch{1.2}
  \footnotesize
  \centering
  \vspace{-1\baselineskip}
  \caption{Quantitative Ablation Study on LE-IRSTD(Mobile Inverted Bottleneck Convolution is marked \textbf{MBConv}, Bottleneck Structure Block is mared \textbf{BSblock}, Adaptive Vision Convolution Stem is mared \textbf{AVCStem}, Ghost Shuffle Convolution is mared \textbf{GSConv})}
  \label{tab:LE-IRSTD-ablation}
  \setlength{\tabcolsep}{6pt}
  \begin{tabular}{c|cccc|cccccc}
      \multirow{2}{*}{Ablation} & \multicolumn{4}{c|}{Module} & \multicolumn{6}{c}{Evaluation Metrics} \\
      & \textbf{MBConv} & \textbf{BSblock} & \textbf{AVCStem} & \textbf{GSConv}  & P $\uparrow$ &  R  $\uparrow$ & F1 $\uparrow$ &  mAP $\uparrow$ & Para $\downarrow$ &  Flops $\downarrow$  \\
  \Xhline{1pt}
  (a) & \xmark & \xmark & \xmark & \xmark &83.5 & 82.4 &  82.8 &  86.4 &  3.0M & 8.2G  \\
  (b) & $\checkmark$ & \xmark & \xmark & \xmark & 88.2  & 84.2 &  86.2 &  89.4& 3.2M & 9.0G  \\
  (c) & $\checkmark$ & $\checkmark$ & \xmark  & \xmark & 88.2  & 81.7 & 85.2 & 90.7  & 2.9M  & 8.4G   \\
  (d) & $\checkmark$ & $\checkmark$ & $\checkmark$ & \xmark  & 89.9  & 84.2 & 86.9 & 92.9 & 2.7M & 7.6G  \\
  \hline
  \rowcolor[rgb]{0.9,0.9,0.9}
  (e) & \textbf{$\checkmark$} & $\checkmark$ & $\checkmark$  & \textbf{$\checkmark$} & \textbf{89.8}  & \textbf{91.2} & \textbf{90.5} & \textbf{95.4} & \textbf{2.6M} & \textbf{7.4G} \\
  
\end{tabular}
\end{table*}

\begin{figure*}
    \centering
    \includegraphics[width=1\linewidth]{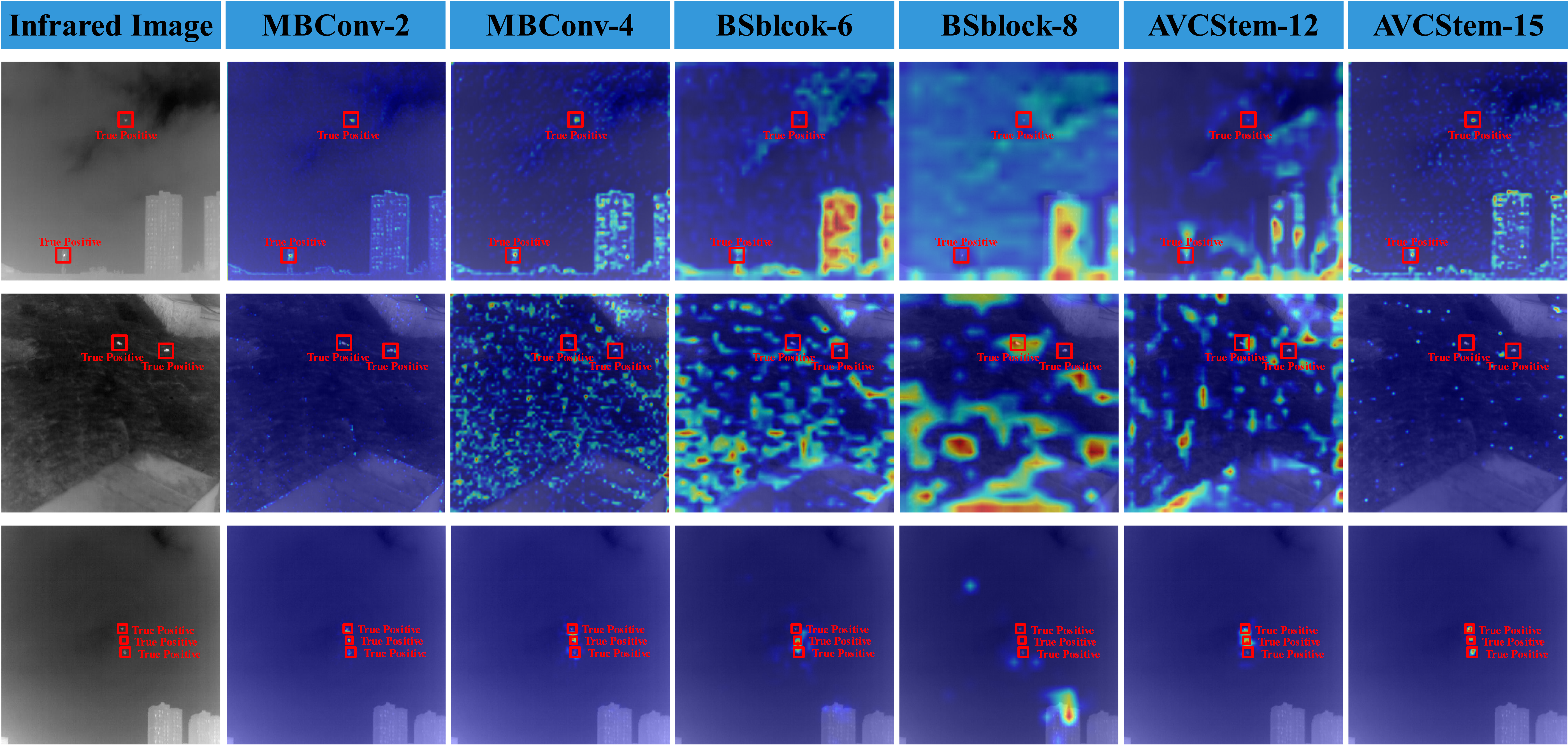}
    \caption{The improved gradmap effects of adding different model components are displayed. The more prominent the gradmap, the stronger the ability to extract attention areas. The red box indicates the true small target, while the other highlighted areas in the image represent interference.}
    \label{fig11:Gradcam}
\end{figure*}

To enhance the interpretability of the model, we visualized the Grad-CAM \cite{selvaraju2017grad} of the added modules to determine whether the modules focus on target regions, as shown in Fig. \ref{fig11:Gradcam}. MBConv-2 indicates the addition of a module in the second stage of YOLOv8-n and its corresponding visualization; other modules follow a similar approach. YOLOv8-n comprises a total of 21 stages, and different model structural layers process features differently, resulting in varying attention areas. However, all ultimately demonstrate a focus on small target regions, indicating that the stacking of modules is effective. Notably, by the 16th stage, the feature attention areas become less prominent, as the output feature representation has already been completed.


\subsubsection{\textbf{Ablation of MBConvblock}} 

For the lightweight feature extraction module, we conducted an ablation study on the model’s width, depth, and input feature map size, as shown in Tab. \ref{tab:MBConv}. To determine the optimal architecture of LE-IRSTD, we replaced the C2f modules in the backbone of YOLOv8-n with MBConvblock and evaluated various parameter configurations. Specifically, width multipliers of 1 and 6 correspond to expansion ratios of $1\times$ and $6\times$ within the MBConvblock. The ablation results demonstrate that when the expansion ratio is greater than 1, an additional expansion convolution is applied before depthwise separable convolution, leading to changes in the convolutional pattern and output channel dimensions; in contrast, an expansion ratio of 1 omits the expansion convolution, performing only depthwise separable convolution. The kernel size was compared between $3\times3$ and $5\times5$ to adjust the receptive field of MBConvblock. For depth adjustment, two C2f modules in the original YOLOv8-n backbone repeated 3 and 6 times, respectively were replaced by MBConv blocks, where two groups of repeated MBConv blocks were used to achieve the corresponding depth. Since multiple bottlenecks in the C2f modules substantially increase computational cost, MBConvblock achieve a favorable trade-off by reducing complexity while maintaining performance. Experimental results indicate that the optimal configuration consists of a $6\times$ expansion ratio, a $3\times3$ convolutional kernel, and depths of 1 and 2.

 
By employing various evaluation metrics for component ablation, experimental results show that the MBConvblock achieved favorable trade-offs between parameter and Flops while significantly improving detection accuracy. This indicates that the adopted MBConvblock effectively optimized the model architecture, achieving efficient resource utilization.


\begin{table}
  \setlength{\abovecaptionskip}{0cm} 
  \renewcommand\arraystretch{1.2}
  \footnotesize
  \centering
  \vspace{-1\baselineskip}
  \caption{Ablation Study on Adjustments to width, depth, and resolution for backbone feature extraction.}
  \label{tab:MBConv}
  \setlength{\tabcolsep}{3pt}
  \begin{tabular}{c|c|c|cccccc}
       \textbf{Width} & \textbf{Convolution} & \textbf{Depth} & P$\uparrow$ & R$\uparrow$ & F1$\uparrow$ & mAP$\uparrow$ & Para$\downarrow$ & Flops$\downarrow$ \\
  \Xhline{1pt}
  1 & 3*3 & 1,2 & 84.5 & 78.9 & 81.6 & 85.4 & 2.3M & 6.0G \\
  2 & 3*3 & 1,2 & 85.8 & 82.6 & 84.1 & 86.2 & 2.3M & 6.5G \\
  3 & 3*3 & 1,2 & 87.6 & 82.5 & 85.0 & 88.9 & 2.4M & 6.62G \\
  4 & 3*3 & 1,2 & 88.3 & 86.1 & 87.2 & 90.4 & 2.5M & 6.9G \\
  5 & 3*3 & 1,2 & 89.7 & 91.2 & 90.5 & 93.4 & 2.5M & 7.2G \\
  \hline
  \rowcolor[rgb]{0.9,0.9,0.9}
  \textbf{6} & \textbf{3*3} & \textbf{1,2}  & \textbf{89.8} & \textbf{91.2} & \textbf{90.5} & \textbf{95.4} & \textbf{2.6M}  &  \textbf{7.4G} \\
  \hline
  6 & 5*5 & 1,2 & 86.1 & 83.2 & 84.6 & 93.7 & 3.5M & 8.8G \\
  6 & 5*5 & 3,6 & 84.3 & 82.5 & 83.4 & 94.3 & 2.8M & 9.1G \\
  
\end{tabular}
\end{table}

\subsubsection{\textbf{Ablation of BSblock}} 


In the C2f module of the YOLOv8 backbone, while the multiple Bottleneck structures enhance the network’s ability to model complex data, the repeated feature transformations do not achieve an ideal lightweight structure. To address this, we introduce the BSblock. Unlike C2f, our approach randomly applies partial convolution to feature maps in the channel dimension. As shown in Tab. \ref{tab:BSblock}, C2f represents the baseline YOLOv8, while BSblock (SC) and BSblock (PConv) denote the BSblock using standard convolution and partial convolution, respectively. The results indicate that the BSblock utilizing PConv outperforms the baseline, and the implementation of partial convolution within the BSblock effectively reduces network parameters.

\begin{table}
  \setlength{\abovecaptionskip}{0cm}  
  \renewcommand\arraystretch{1.2}
  \footnotesize
  \centering
  \vspace{-1\baselineskip}
  \caption{Ablation Study on comparing partial convolutions in BSblock with standard convolutions.}
  \label{tab:BSblock}
 \setlength{\tabcolsep}{5pt}
  \begin{tabular}{c|cccccc}
       \textbf{Ablation} & P$\uparrow$ & R$\uparrow$ & F1$\uparrow$ & mAP $\uparrow$ & Para$\downarrow$ & Flops$\downarrow$ \\
  \Xhline{1pt}
  C2f & 83.5 & 82.4 & 82.8 & 86.4 & 3.0M & 8.2G  \\
  BSblock(SC) & 87.2 & 87.3 & 89.9 & 90.9 & 3.3M & 8.5G  \\
  \hline
  \rowcolor[rgb]{0.9,0.9,0.9}
   \textbf{BSblock(PConv)} &  \textbf{89.8} & \textbf{91.2}  & \textbf{90.5} & \textbf{95.4} & \textbf{2.6M}  &  \textbf{7.4G}   \\
  
\end{tabular}
\end{table}

\subsubsection{\textbf{Ablation of AVCStem}} 

For neck feature fusion, we employ the AVCStem module, which performs parallel processing on spatial and channel dimensions to enhance detection efficiency. Additionally, we have optimized the subsequent convolution blocks. Standard convolutional kernels typically use fixed square shapes to extract features, often overlooking the fact that small targets in infrared images can vary significantly in shape. This fixed-size square kernel may struggle to accurately define small targets, especially in areas with irregular edges. In contrast, VKConv modifies the convolution shapes based on standard convolution to better accommodate the diverse forms of features, thereby enhancing feature information and facilitating better resolution for subsequent detection heads. As shown in Tab. \ref{tab:AVCStem}, SC represents the standard convolution block, while VKConv denotes the deformable convolution. Using standard convolution results in a higher computational load and lower detection accuracy compared to VKConv.

\begin{table}
  \setlength{\abovecaptionskip}{0cm}  
  \renewcommand\arraystretch{1.2}
  \footnotesize
  \centering
  \vspace{-1\baselineskip}
  \caption{Ablation Study on comparing variable convolutions with standard convolutions in AVCStem.}
  \label{tab:AVCStem}
 \setlength {\tabcolsep}{5pt}
  \begin{tabular}{c|cccccc}
       \textbf{Ablation} &P$\uparrow$ & R$\uparrow$ & F1$\uparrow$ & mAP$\uparrow$ & Para$\downarrow$ & Flops$\downarrow$\\
  \Xhline{1pt}
  SC & 83.37 & 85.75 & 82.77 & 86.78 & 3.51M & 8.86G  \\
  \hline
  \rowcolor[rgb]{0.9,0.9,0.9}
   \textbf{VKConv} &  \textbf{89.8} & \textbf{91.2}  & \textbf{90.8} & \textbf{95.4} & \textbf{2.6M}  &  \textbf{7.4G}   \\
  
\end{tabular}
\end{table}

\subsubsection{\textbf{Ablation of GSconv}} 

The albation of GSConv is shown in Tab. \ref{tab:GSconv}:
SC + DWConv represents the combination of standard and depthwise convolution.
SC + shuffle indicates that channel shuffle is performed directly after standard convolution.
SC + DWConv + shuffle means that after standard convolution, depthwise convolution is applied, followed by channel shuffle.
The results shows that the combination of SC + shuffle results in a significant increase in Flops. The Flops calculation is based on the product of the feature map size, the convolution kernel, and the input/output channels. When depthwise convolution is not applied, the feature map size is larger, leading to an increase in Flops during shuffling. The final combination of SC + DWConv + shuffle demonstrates that our GSConv not only improves detection accuracy but also reduces the number of parameter and computational load.

\begin{table}
  \setlength{\abovecaptionskip}{0cm}  
  \renewcommand\arraystretch{1.2}
  \footnotesize
  \centering
  \vspace{-1\baselineskip}
  \caption{Ablation Study on standard convolutions, depthwise convolutions, and shuffle in the GSConv architecture.}
  \label{tab:GSconv}
 \setlength{\tabcolsep}{4pt}
  \begin{tabular}{c|cccccc}
       \textbf{Ablation} &P$\uparrow$ & R$\uparrow$ & F1$\uparrow$ & mAP$\uparrow$ & Para$\downarrow$ & Flops$\downarrow$ \\
  \Xhline{1pt}
  SC+DWConv & 84.9 & 75.3 & 79.8 & 85.9 & 3.5M & 8.9G  \\
  SC+shuffle & 86.2 & 83.9 & 85.1 & 89.9 & 3.8M & 9.3G  \\
  \hline
  \rowcolor[rgb]{0.9,0.9,0.9}
   \textbf{SC+DWConv+shuffle} &  \textbf{89.8} & \textbf{91.2}  & \textbf{90.5} & \textbf{95.4} & \textbf{2.6M}  &  \textbf{7.4G}   \\
   
\end{tabular}
\end{table}

\subsubsection{\textbf{Ablation of loss function weights parameter.}} 

An ablation study was conducted on the three loss weighting parameters $\lambda_{1}$, $\lambda_{2}$ and $\lambda_{3}$ in YOLOv8, which correspond to the classification loss (cls), Regression Loss (CIoU+DFL), respectively. The experimental results shown in Tab. \ref{tab:loss} demonstrate that the optimal detection precision and accuracy are achieved when $\lambda_{1}$ = 0.02, $\lambda_{2}$ = 0.02, $\lambda_{3}$ = 0.49. This can be attributed to the nature of the task, which involves only a single target category; assigning a high weight to the classification loss may introduce redundant optimization noise and hinder model convergence. In contrast, increasing the weights of CIoU and DFL both closely related to object localization-significantly improves the quality of bounding box regression. Therefore, reducing the contribution of cls while emphasizing CIoU and DFL leads to enhanced localization accuracy and overall detection performance.

\begin{table}[htbp]
  \setlength{\abovecaptionskip}{0cm}
  \renewcommand\arraystretch{1.2}
  \footnotesize
  \centering
  \caption{Ablation study on loss function weights parameter.}
  \label{tab:loss}
  \setlength{\tabcolsep}{6pt}  
  \begin{tabular}{ccc|cccc}
    \textbf{$\lambda_{1}$} & \textbf{$\lambda_{2}$} & \textbf{$\lambda_{3}$} & P$\uparrow$ & R$\uparrow$ & F1$\uparrow$ & mAP$\uparrow$ \\
    \hline
    0.8 & 0.1 & 0.1 & 82.9 & 83.4 & 83.1 & 86.7 \\
    0.7 & 0.15 & 0.15 & 83.1 & 84.7 & 83.9 & 87.0 \\
    0.6 & 0.2 & 0.2 & 83.6 & 84.7 & 84.1 & 87.2 \\
    0.5 & 0.25 & 0.25 & 83.7 & 85.7 & 84.7 & 87.8 \\
    0.4 & 0.3 & 0.3 & 85.0 & 86.7 & 85.8 & 87.7 \\
    0.3 & 0.35 & 0.35 & 85.1 & 87.2 & 86.2 & 88.1 \\
    0.2 & 0.4 & 0.4 & 85.9 & 87.9 & 86.9 & 89.8 \\
    0.1 & 0.45 & 0.45 & 86.3 & 88.5 & 87.4 & 90.7 \\
    0.08 & 0.46 & 0.46 & 86.5 & 88.6 & 87.5 & 82.3 \\
    0.06 & 0.47 & 0.47 & 89.4 & 89.4 & 89.4 & 93.3 \\
    0.04 & 0.48 & 0.48 & 89.2 & 89.8 & 89.5 & 94.1 \\
    \hline
    \rowcolor[rgb]{0.9,0.9,0.9}
    \textbf{0.02} & \textbf{0.49} & \textbf{0.49} & \textbf{89.8} & \textbf{91.2} & \textbf{90.5} & \textbf{95.4} \\
  \end{tabular}
\end{table}


\section{Conclusion} \label{sec:conclusion}
This paper proposed the LE-IRSTD, which targets the challenges of model lightweighting and practical performance that has been overlooked in current advancements of the YOLO series. Specifically, we enhanced the YOLOv8-n backbone network, where the MBConvblock employs compound scaling to address the repeated stacking of feature extraction modules, enabling the model to achieve higher detection accuracy through a well-designed architecture. The BS block, applied in the deep feature extraction stage, reduces model parameters while simultaneously enhancing detection accuracy. The AVCStem in the neck facilitates dual-channel detection information and uses VKConv to flexibly extract diverse feature information. Lastly, GSConv is employed to mitigate feature loss during downsampling. In comparative experiments, we evaluated our method against various traditional and deep learning approaches, and the results demonstrate that our method meets the requirements for model lightweighting and improved detection accuracy, outperforming other advanced methods.



\bibliographystyle{IEEEtran}
\bibliography{./reference.bib}

\end{document}